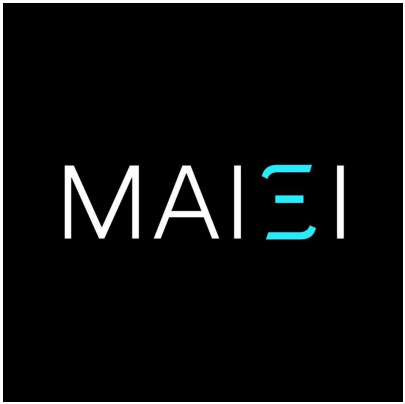

Montreal AI Ethics Institute

*An international, non-profit research institute helping humanity define its place in a world increasingly driven and characterized by algorithms*

Website: https://montrealethics.ai
Newsletter: https://aiethics.substack.com

# Montreal AI Ethics Institute's (MAIEI) Submission to the World Intellectual Property Organization (WIPO) Conversation on Intellectual Property (IP) and Artificial Intelligence (AI) Second Session

## Theme: IP Protection for AI-Generated and AI-Assisted Works

Based on insights from the Montreal AI Ethics Institute (MAIEI) staff and supplemented by workshop contributions from the AI Ethics community convened by MAIEI on July 5, 2020.

<u>Primary contacts for the report:</u>

Allison Cohen (allison@montrealethics.ai)
AI Ethics Researcher, Montreal AI Ethics Institute

Abhishek Gupta (abhishek@montrealethics.ai)
Founder, Montreal AI Ethics Institute
Machine Learning Engineer, Microsoft

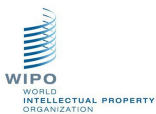

WIPO Conversation on IP and AI

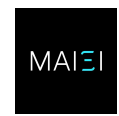

Montreal AI Ethics Institute

# Introduction

This document posits that, at best, a tenuous case can be made for providing AI exclusive IP over their "inventions". Furthermore, IP protections for AI are unlikely to confer the benefit of ensuring regulatory compliance. Rather, IP protections for AI "inventors" present a host of negative externalities and obscures the fact that the genuine inventor, deserving of IP, is the human agent. This document will conclude by recommending strategies for WIPO to bring IP law into the 21$^{st}$ century, enabling it to productively account for AI "inventions".

# Tenuous Benefits of AI Innovators

As is the case for most policy tools, IP laws have been designed to incentivize and disincentivize human behavior to achieve an "optimal" goal. The goal of the patent system is to motivate people to invest time, money, and energy into creating novel and meaningful inventions for society.

However, AI technology, in and of itself, is impervious to behavioral nudging efforts by policy tools because, unlike humans, AI is not *knowingly* self-interested. This reality begs the question: even in the best-case scenario, what would IP for AI achieve? Why should the definition of IP be broadened to account for AI "inventors" when those "inventors" will not change their behavior in light of these protections? Since AI does not actively: i) invest in R&D; ii) decide to develop innovative rather than generic approaches to problem-solving; nor does it, iii) understand the concept of "meaningful to our society", then what role will IP protections for AI play in promoting innovation?

A useful analogy, which demonstrates the inapplicability of legal frameworks to AI systems, is the concept of applying criminal law to self-driving cars. Unlike human drivers, self-driving cars are not incentivized to drive safely because of legal penalties or demerit points. In fact, besides being a sequence of 1s and 0s, self-driving cars have no idea what the concept of "legal penalties" or "demerit points" are; let alone feeling motivated to behave according to the incentive structure they've established. Therefore, developers of self-driving cars are not requesting legal penalties and demerit points be expanded to ensure safe driving by AI technology; instead, developers are focused on the technical components that are compatible with this technology to ensure it is safe. Why then are we attempting to expand IP law to influence AI technology when those incentives are not compatible with the technology's behavior?

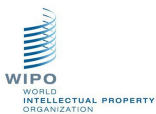
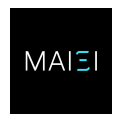

WIPO Conversation on IP and AI　　　　　　　　　　　　　　　　　Montreal AI Ethics Institute

Expanding legal frameworks to govern AI "actors" is prefaced on false assumptions and might, in fact, disincentivize genuine creative and inventive efforts by human actors.

# IP as an Ineffective Regulatory Compliance Mechanism for AI Innovators

Some WIPO interveners may argue that, when applied to AI, IP can become a useful tool for ensuring regulatory compliance. But, ironically, IP can actively undermine AI regulation by taking problematic features of the technology, such as a lack of transparency and explainability[1], and mis-interpreting them as valuable characteristics like creativity and non-obviousness, which make the AI worthy of IP protection.

Currently, in an attempt to defend a patent, human inventors must disclose the method and apparatus used to arrive at their invention. This process results in human agents receiving credit for their invention but also liability for that which they've created. However, if IP rights were conferred onto AI, how would the technology disclose its own method and apparatus when deep learning is inherently a black box? Furthermore, if AI can take all the credit then with whom does the responsibility lie? It is possible that companies profiting from the technology's invention could simply wipe their hands clean of liability as the AI's "creativity" has been legitimized by the IP and need not be explained for fear of disproving "non-obviousness" or "inventiveness". Therefore, rather than holding the AI to a higher standard of responsibility, the IP protection would turn features we've been critical of into characteristics worthy of praise and protection.

Our community members believe that Uber settled its legal dispute with a woman killed by their self-driving cars[2] for this exact reason. Had Uber gone to court, their data and model would have been disclosed and the flaws in the programming would need to be addressed. However, Uber does not know where the flaw in the programming lies and nor do their lawyers. As legal liability can clearly not be traced back to the algorithm, this AI cannot be deemed to be the "inventor".

Furthermore, those AI technologies that egregiously violate regulations are unlikely to file a patent application. For example, it is improbable that a problematic deep fake, which takes the face of a high-profile individual and transplants it onto a pornographic image[3], will claim IP protection. Therefore, IP protection will not necessarily be an effective regulatory compliance mechanism as those inventions in breach of the law, and in need of regulation, would be unlikely to claim IP rights.

---

[1] Adadi, A., & Berrada, M. (2018). Peeking inside the black-box: A survey on Explainable Artificial Intelligence (XAI). IEEE Access, 6, 52138-52160.
[2] Kohli, P., & Chadha, A. (2019, March). Enabling pedestrian safety using computer vision techniques: A case study of the 2018 uber inc. self-driving car crash. In Future of Information and Communication Conference (pp. 261-279). Springer, Cham.

## Negative Externalities of IP for AI Innovators

While the benefits of IP for AI are tenuous, at best, the negative externalities are severe, at worst. To begin, conferring IP rights onto AI technology may profoundly impact the perception of this technology in society. Once rights that are typically granted to humans are granted to AI technologies, it is likely that these algorithms will be viewed as equally valuable when compared to human creators and worthy of the same rights. This assertion may catalyze the realization of a dystopian society wherein human values and freedoms are valued evenly with AI's "values" and "freedoms", with unspeakable implications for our democracy, autonomy and economic opportunity.

According to Abeba Birhane and Jelle van Dijk, authors of "Robot Rights? Let's talk about Human Welfare Instead"[4], the debate about the rights of robots is coming at the expense of urgent ethical concerns including machine bias, machine elicited human labor exploitation, and the erosion of privacy, all of which are impacting the most vulnerable in our society. The authors argue that the most pressing ethical discussion in AI is not whether robots are entitled to rights but rather the utter lack of responsibility in designing, selling and deploying such machines, with detrimental implications for the lives of human beings.

Therefore, the negative externalities that may arise as a result of conferring IP rights to AI seem to outweigh the benefits, to say the least.

## Humans Drive Value in AI Inventions

Fundamentally, AI "innovation" is created with human time, energy and money and the invention is only valuable in the sense that it is valued by humans. Since the AI's accuracy depends on quality data, obtained by the human, and since the AI's design depends on that which the society deems relevant, identified by the human, it is the human creator rather than the AI that confers value with regards to the AI's creation. Therefore, those responsible for the AI technology's development should own the technology's IP rights.

---

[3] Harris, D. (2018). Deepfakes: False pornography is here and the law cannot protect you. Duke L. & Tech. Rev., 17, 99.
[4] Birhane, A., & van Dijk, J. (2020, February). Robot Rights? Let's Talk about Human Welfare Instead. In Proceedings of the AAAI/ACM Conference on AI, Ethics, and Society (pp. 207-213).

Furthermore, while AI algorithms exist mostly in the public domain, the uniqueness of AI inventions is derived from the system's pipeline arrangement. It is human ingenuity that organizes the elements of AI into a pipeline in order for the technology to provide results that are new and unique. For example, the pipeline arrangement generates parameters obtained from training on the input dataset and the hyperparameters obtained from tuning the algorithm to optimize for some learning objective. Therefore, AI IP rights should be conferred onto humans rather than the AI itself as it is the humans that design that which is perceived as AI's "inventiveness".

Thus, while AlphaGo's inventive computer program ostensibly trained itself using reinforcement learning to become the best Go player in the world[5], the system itself was created by engineers. There was no consciousness or creative intervention, it was purely mathematical equations derived from AlphaGo's initial programming. As a result, rather than comparing the AI to the human inventor it would be more appropriate to compare AI to the microscope used by the human inventor in pursuit of the human's own creation.

# Bringing IP Law in the 21st Century

IP law still has a role to play in the AI space. IP rights can prevent the abuse of trade secrets, enhance the flow of information, and even incentivize remuneration for the data of individual consumers. However, to confer these benefits, IP protections must be applied in new ways.

In order to prevent the use of trade secrets and enhance the flow of information, it is recommended that either companies assign inventor status to the human creator(s) or create a new category for IP law wherein the AI system doesn't own the patent rights but the patent cannot be copied without consequence.

If data were patentable, obtaining that data could be more significantly budgeted into R&D costs. This would incentivize higher quality data retrieval and potential remuneration for individuals in exchange for their data. This practice could enhance fairness in the AI ecosystem and increase representation in the datasets.

---

[5] Borowiec, S. (2016). AlphaGo seals 4-1 victory over Go grandmaster Lee Sedol. The Guardian, 15.

In terms of data disclosure, a summary of aggregate statistics could be disclosed through the use of: datasheets for datasets[6], nutrition labels for datasets[7], factsheets for AI systems[8] or data dictionaries[9]. Each of these mechanisms will reveal the dataset at a level that is not too precise while still allowing for patentable IP.

# Conclusion

Although IP protections for AI inventors might seem promising on the surface, it remains unclear whether a genuine case can be made to demonstrate the benefit of this policy; while the consequences of this policy are clear. These IP protections are unlikely to be effective regulatory mechanisms, they set a dangerous precedent for "robot rights" and they remove agency and attribution of real people who deserve credit for those creations. However, IP law *can* promote responsible AI inventions by incentivizing humans to exchange information and beginning a conversation about remuneration in exchange for data, both of which are a welcome advent to AI in the 21st century.

---

[6] Gebru, T., Morgenstern, J., Vecchione, B., Vaughan, J. W., Wallach, H., Daumé III, H., & Crawford, K. (2018). Datasheets for datasets. arXiv preprint arXiv:1803.09010.
[7] Holland, S., Hosny, A., Newman, S., Joseph, J., & Chmielinski, K. (2018). The dataset nutrition label: A framework to drive higher data quality standards. arXiv preprint arXiv:1805.03677.
[8] Arnold, M., Bellamy, R. K., Hind, M., Houde, S., Mehta, S., Mojsilović, A., ... & Reimer, D. (2019). FactSheets: Increasing trust in AI services through supplier's declarations of conformity. IBM Journal of Research and Development, 63(4/5), 6-1.
[9] Navathe, S. B., & Kerschberg, L. (1986). Role of data dictionaries in information resource management. Information & management, 10(1), 21-46.